\newif\ifconfver
\confverfalse      %declaring conference version false
\ifconfver
	\documentclass{article}
	\usepackage{spconf}
	\title{Divide and Conquer: One-Bit MIMO-OFDM Detection by Inexact Expectation Maximization}
	\name{Mingjie Shao and Wing-Kin Ma
		 \thanks{\scriptsize This work was supported by a General Research Fund (GRF) of the Hong Kong Research Grant Council under Project ID CUHK 142017318.}
}

	\address{
 \normalsize Department of Electronic Engineering, The Chinese University of Hong Kong,
    Hong Kong SAR of China \\
}
\else
	\documentclass[11pt]{article}
	\usepackage{fullpage}
    \title{Divide and Conquer: One-Bit MIMO-OFDM Detection by Inexact Expectation Maximization}
	\author{
    Mingjie Shao and Wing-Kin Ma\\~ \\
    Department of Electronic Engineering, The Chinese University of Hong Kong, \\
    Hong Kong SAR of China \\ ~ \\
    E-mails: mjshao@ee.cuhk.edu.hk, wkma@ee.cuhk.edu.hk
    }
\fi
\usepackage{amsmath,graphicx}

\usepackage{calc,amsfonts,amssymb,bm,url,color,theorem,cite}
\usepackage{psfrag,float}
\usepackage{algorithm}
\usepackage{algorithmic}
\usepackage{mathtools,lipsum,cuted,multicol}
\usepackage{filecontents}
\usepackage{subcaption,epstopdf}

\usepackage{bbold}
\definecolor{orange}{RGB}{255,107,0}

\newtheorem{Fact}{Fact}

\theorembodyfont{\rmfamily}

\newtheorem{Remark}{Remark}

\newcommand\ba{\ensuremath{{\bm a}}}

\newcommand\bs{\ensuremath{{\bm s}}}

\newcommand\bw{\ensuremath{{\bm w}}}

\newcommand\bh{\ensuremath{{\bm h}}}

\newcommand\jj{\ensuremath{{\frak j}}}

\newcommand{\setD}{\mathcal{D}}
\newcommand{\setX}{\mathcal{X}}

\newcommand{\Rbb}{\mathbb{R}}
\newcommand{\Cbb}{\mathbb{C}}

\newcommand{\Exp}{\mathbb{E}}

\newcommand\Ree{\ensuremath{{\rm Re}}}
\newcommand\Imm{\ensuremath{{\rm Im}}}

\newcommand\bx{\ensuremath{{\bm x}}}

\newcommand\bH{\ensuremath{{\bm H}}}

\newcommand\bS{\ensuremath{{\bm S}}}

\newcommand\bF{\ensuremath{{\bm F}}}

\newcommand{\setU}{\mathcal{U}}

\newcommand{\setS}{\mathcal{S}}

\newcommand\br{\ensuremath{{\bm r}}}

\makeatletter
\def\bstctlcite{\@ifnextchar[{\@bstctlcite}{\@bstctlcite[@auxout]}}
\def\@bstctlcite[#1]#2{\@bsphack
  \@for\@citeb:=#2\do{%
    \edef\@citeb{\expandafter\@firstofone\@citeb}%
    \if@filesw\immediate\write\csname #1\endcsname{\string\citation{\@citeb}}\fi}%
  \@esphack}
\makeatother
\begin{document}
%\ninept

\ifconfver
	\ninept
\fi

\bstctlcite{IEEEexample:BSTcontrol}
\maketitle
\begin{abstract}
Adopting one-bit analog-to-digital convertors (ADCs) for massive multiple-input multiple-output (MIMO) implementations has great potential in reducing the hardware cost and power consumption.
However, distortions caused by quantization raise great challenges.
In MIMO orthogonal frequency-division modulation (OFDM) detection, coarse quantization renders the orthogonal separation among subcarriers inapplicable, forcing us to deal with a problem that has a very large problem size.
In this paper we study the expectation-maximization (EM) approach for one-bit MIMO-OFDM detection.
The idea is to iteratively decouple the MIMO-OFDM detection problem among subcarriers.
Using the perspective of block coordinate descent, we describe inexact variants of the classical EM method for providing more flexible and computationally efficient designs.
Simulation results are provided to illustrate the potential of the divide-and-conquer strategy enabled by EM.
\end{abstract}
%
% \begin{keywords}
%one-bit ADCs, MIMO-OFDM, maximum-likelihood detection, expectation-minimization
% \end{keywords}

%
\section{Introduction}

The use of low-resolution analog-to-digital convertors (ADCs), particularly one-bit ADCs, provides a promising candidate for power-efficient and cost-reduced implementation of massive multiple-input multiple-output (MIMO) systems \cite{mezghani2010multiple,risi2014massive,juncil2015near,wen2015bayes,hong2017weighted,MollenCLH17,JacobssonDCGS17}.
In uplink MIMO, the  number of ADCs scales with the number of antennas at the base station (BS),
and using high-resolution ADCs in massive MIMO means significant rise in price and power consumption.
Moreover, the use of one-bit ADCs is able to significantly simplify the designs of radio-frequency (RF) chains and automatic gain control.
However, one-bit quantization  incurs severe distortion on the signals, which necessitates effective signal processing techniques to combat the quantization effects.

Recently, it has been shown that given a massive number of antennas at the BS, the  quantization effects of  one-bit ADCs can be mitigated \cite{juncil2015near,LiTSMSL17,shao2020binary}.
One-bit massive MIMO systems have spurred immense research interest in aspects such as  asymptotic system performance analyses and  efficient channel estimation/data detection.
%While linear detectors perform well under ADCs with moderate resolution, e.g., $3\sim 6$-bit, they can suffer from severe performance degradation when one-bit ADCs are considered \cite{risi2014massive,studer2016quantized}.
%Advanced designs have achieved better performance.
For instance, the performances of
linear estimators and detectors have been
%extensively
studied in \cite{risi2014massive,MollenCLH17,JacobssonDCGS17};
%, though the performances of such linear designs are usually inferior to those of the advanced (non-linear) designs discussed below;
maximum-likelihood (ML) data detection has been considered in \cite{juncil2015near,shao2020binary};
%, where sphere relaxation was used in \cite{juncil2015near} and homotopy and deep-learning based approach was explored in \cite{shao2020binary};
approximate message passing algorithms for data detection have been studied in \cite{wang2014multiuser,wen2015bayes};
%for Rayleigh faded channels;
the $\Sigma\Delta$ ADC architecture has been used in \cite{rao2020massive}.
% for serving restricted angle sectors.

The majority of the current one-bit MIMO studies focus on  frequency-flat  channels.
In comparison, there is a paucity of studies on one-bit MIMO with orthogonal frequency division multiplexing (OFDM) and over frequency-selective  channels \cite{MollenCLH17,studer2016quantized,stockle2016channel,plabst2018efficient, Mirfarshbafan2020}.
This paper studies MIMO-OFDM detection. The problem  is, in essence, a large-scale one;
the size is the number of users times the subcarrier size (which is of the order of hundreds or thousands).
In classical, or unquantized, MIMO-OFDM detection,
the problem can be orthogonally decoupled among subcarriers;
detection can then be performed independently for each subcarrier.
In the one-bit case, however, the orthogonality among subcarriers is destroyed by the  quantization.
This prevents us from taking advantage of the subcarrier-wise orthogonal separation---at least not directly.
The existing one-bit MIMO-OFDM detection studies can be taxonomized into two classes:
i) applying a first-order method, such as the projected gradient (PG) method, to the large-scale detection problem  \cite{studer2016quantized,Mirfarshbafan2020}; and
ii) using expectation-maximization (EM)  \cite{stockle2016channel,plabst2018efficient}.
%Still, the computational complexities of these approaches can be large if the number of subcarriersis large.
In particular, the M-step of EM allows us to decouple the
one-bit MIMO-OFDM detection problem into a number of subcarrier-wise independent problems---just like the unquantized MIMO-OFDM detection.
Curiously, the potential of divide-and-conquer enabled by EM has not drawn much attention in the literature.

%Is it possible that 1) the one-bit ML MIMO-OFDM detection problem can be decoupled among subcarriers?
%2) the subproblem in each subcarrier can be efficiently handled by

In this paper, we examine the EM method for one-bit ML MIMO-OFDM detection, with emphasis on designing efficient algorithms and providing insight of the relation with unquantized MIMO-OFDM.
%It is desirable that if we can decouple the one-bit MIMO-OFDM detection problem among the subcarriers in the algorithmic development, which allows us to take advantage of parallel processing for acceleration.
%%Moreover, it can be highly appreciated if we can build relationship of one-bit MIMO-OFDM detection with classical MIMO-OFDM detection with full-resolution
%In this paper, we present such a solution.
%We study the ML detection formulation for one-bit MIMO-OFDM system and develop an expectation-maximization (EM) approach to handle the resulting problem.
%In the E-step, we construct a  tight upper bound of the negative log-likelihood function; the E-step admits a closed-form solution.
%The upper bound is then minimized in the M-step.
We interpret EM from a block coordinate descent (BCD) and variational viewpoint.
Such viewpoint not only provides an alternative interpretation of the classical EM, it also leads to inexact EM variants for better computational efficiency.
%recover the classical EM algorithm, but also generate new variants with inexact update rule in the M-step to reduce computational complexity.
%Two salient features of the EM method are worth noting.
%First, the M-step can be decoupled among subcarriers and can be solved in parallel, just as in classical ML MIMO-OFDM detection.
%Second, in each subcarrier, the detection problem amounts to a classical ML MIMO detection problem, where off-the-shelf proverbial classical MIMO detectors
%can be readily applied.
%Thus, the overall EM algorithm can be computationally efficient.
%The EM method reveals an important revelation:
%{\it The one-bit ML MIMO-OFDM detection problem can be solved in almost the same way as unquantized ML MIMO-OFDM detection problem, aided by an additional E-step.}
Simulation results show that our proposed inexact EM exhibits satisfactory bit-error rate performance with reasonable  computational complexity.
%it does so with reasonable light computational complexity.

\section{System Model}

\subsection{Unquantized MIMO-OFDM}

We first review the classical scenario of unquantized MIMO-OFDM detection to provide the reader with the context or to recall the reader the basics.
Consider an uplink scenario wherein a number of $K$ single-antenna users simultaneously transmit blocks of symbols to an $N$-antenna base station (BS) over a frequency-selective channel.
The modulation scheme is OFDM.
Under some standard assumptions (such as cyclic prefix insertion at the transmitter side and guard time removal at the receiver side),
the received signal over one transmission block can be described by the following formula
\begin{equation}
\bm r_n = \textstyle \sum_{k=1}^K \bm H_{n,k} \bm F^H \bm s_k + \bm v_n,
\qquad n=1,\ldots,N,
\end{equation}
where
$\bm r_n \in \mathbb{C}^W$ is the time-domain received signal block at the $n$th antenna of the BS;
$W$ is the block length;
$\bm H_{n,k} \in \mathbb{C}^{W \times W}$ is a circulant matrix whose coefficients are the time-domain impulse response of the channel from user $k$ to the $n$th antenna of the BS;
$\bm F \in \mathbb{C}^{W \times W}$ is the (unitary) $W$-point DFT matrix;
$\bm s_k \in \mathbb{C}^W$ is the symbol block transmitted by user $k$;
$\bm v_n$ is i.i.d. circular Gaussian noise with mean zero and variance $\sigma_C^2$.
By the eigendecomposition $\bm H_{n,k} = \bF^H \bm D_{n,k} \bF$, where $\bm D_{n,k} = {\rm Diag}(\check{\bm h}_{n,k})$ and $\check{\bm h}_{n,k}$ is the DFT of the time-domain channel impulse response, we can write
\begin{equation}
\bm r_n = \textstyle \bm F^H ( \sum_{k=1}^K  ( \check{\bm h}_{n,k} \odot \bm s_k )) + \bm v_n,
\end{equation}
where $\odot$ denotes the Hadamard product.

The problem is to detect $\bm S = [~ \bs_1,\ldots,\bs_K ~]$ from $\bm r_1,\ldots,\bm r_N$, given knowledge of the channel.
We consider the maximum-likelihood (ML) detector, which is known to be given by
\begin{equation} \label{eq:ml1}
\min_{\bm S \in \mathcal{S}^{W \times K} } \textstyle \sum_{n=1}^N \| \bm r_n - \bm F^H ( \sum_{k=1}^K  ( \check{\bm h}_{n,k} \odot \bm s_k )) \|^2,
\end{equation}
where $\mathcal{S}$ denotes the symbol constellation.
The ML problem~\eqref{eq:ml1} is a large-scale problem, having $WK$ unknowns (note that $W$ is large, typically from $128$ to a few thousands).
However, it can be decoupled into a number of $W$ problems.
By noting $\| \bm r_n - \bm F^H ( \sum_{k=1}^K  ( \check{\bm h}_{n,k} \odot \bm s_k )) \|^2 = \| \bm F \bm r_n - \sum_{k=1}^K  ( \check{\bm h}_{n,k} \odot \bm s_k ) \|^2$,
we can equivalently rewrite the objective function of \eqref{eq:ml1} as
\begin{equation}
\textstyle \sum_{w=1}^W \| \check{\bm r}_w - \bm H_w \check{\bm s}_w \|^2.
\end{equation}
Here, $\check{\bm r}_w = (\tilde{r}_{1,w},\ldots, \tilde{r}_{N,w} )$, with $\tilde{\bm r}_n = \bm F \bm r_n$, describes the received signal at subcarrier $w$;
$\bm H_w = [\check{ h}_{n,k,w} ]_{n,k} \in \mathbb{C}^{N \times K}$ is the MIMO channel frequency response at subcarrier $w$;
$\check{\bm s}_w = ( s_{1,w},\ldots,s_{K,w} )$ is the multiuser symbol vector at subcarrier $w$.
Hence, we can decouple the ML problem \eqref{eq:ml1} into
\begin{equation}\label{eq:ml3}
\min_{ \check{\bm s}_w \in \mathcal{S}^K } \| \check{\bm r}_w - \bm H_w \check{\bm s}_w \|^2, \qquad w=1,\ldots,W,
\end{equation}
which can be handled independently and have a much smaller size.

\subsection{Quantized MIMO-OFDM}

Now, consider the same scenario as above, but with one-bit quantized measurements
\[
\bm y_n = Q( \bm r_n ),
\]
where $Q$ applies the in-phase quadrature-phase (IQ) quantization $Q(x)= {\rm sgn}(\Ree(x)) + \jj \cdot  {\rm sgn}(\Imm(x))$ in the element-wise fashion.
The ML detector in this case is given by
\begin{equation} \label{eq:ml2}
\min_{\bm S \in \mathcal{S}^{W \times K} } \textstyle F(\bm S):= \sum_{n=1}^N \sum_{w=1}^W f_{n,w}(\bm S),
\end{equation}
where
\begin{align*}
f_{n,w}(\bm S) =& -\log \Phi \left( \frac{\Ree(y_{n,w}) \Ree(z_{n,w})}{\sigma} \right) \\
 & ~~ - \log \Phi \left( \frac{\Imm(y_{n,w}) \Imm(z_{n,w})}{\sigma} \right), \\
\bm z_n & = \textstyle \bm F^H ( \sum_{k=1}^K  ( \check{\bm h}_{n,k} \odot \bm s_k )),
\end{align*}
%
%\begin{align*}
%f_{n,w}(\bm S) & =f_{n,w}^R (\bm S) + f_{n,w}^I (\bm S), \\
%f_{n,w}^R (\bm S) & = -\log \Phi \left( \frac{\Ree(y_{n,w}) \Ree(z_{n,w})}{\sigma} \right), \\
%f_{n,w}^I (\bm S) & = - \log \Phi \left( \frac{\Imm(y_{n,w}) \Imm(z_{n,w})}{\sigma} \right), \\
%\bm z_n & = \textstyle \bm F^H ( \sum_{k=1}^K  ( \check{\bm h}_{n,k} \odot \bm s_k )),
%\end{align*}
%\begin{align*}
%f_{n,w}(\bm S) & =f_{n,w}^R (\bm S) + f_{n,w}^I (\bm S), \\
%f_{n,w}^R (\bm S) & = -\log \Phi \left( \frac{\Ree(y_{n,w}) \Ree(z_{n,w})}{\sigma} \right), \\
%f_{n,w}^I (\bm S) & = - \log \Phi \left( \frac{\Imm(y_{n,w}) \Imm(z_{n,w})}{\sigma} \right), \\
%\bm z_n & = \textstyle \bm F^H ( \sum_{k=1}^K  ( \check{\bm h}_{n,k} \odot \bm s_k )),
%\end{align*}
with $\Phi(x) = \frac{1}{\sqrt{2\pi}} \int^x_{-\infty} e^{-t^2/2} dt$ and $\sigma^2 = \sigma_C^2/2$.
One will find that the decoupling trick we saw in the preceding subsection does not apply to the one-bit ML problem \eqref{eq:ml2}.
Or, to describe in an intuitive way, one-bit quantization destroys the orthogonality of OFDM.
%The main challenge with the ML problem \eqref{eq:ml2} and the approximations thereof lies in the large problem size.
As a result, the large problem size of the ML problem \eqref{eq:ml2} serves as the main challenge.

We notice only a few works that deal with the one-bit MIMO-OFDM detection problem \eqref{eq:ml2} or its multi-bit quantized counterpart.
In \cite{Mirfarshbafan2020}, the authors consider a convex relaxation of \eqref{eq:ml2}.
%They implement the convex relaxation by the FISTA algorithm [X], which is reminiscent of a gradient descent scheme.
They implement the convex relaxation by the projected gradient algorithm.
Given the large problem scale, computing the gradient takes a non-negligible amount of computations.
We can exploit the problem structure related to DFT (or FFT) to reduce the computational cost, but still the cost is not cheap.
In \cite{plabst2018efficient}, the authors consider an approximation of \eqref{eq:ml2} that can be seen as  unconstrained relaxation.
There they apply expectation maximization (EM)---in which the M-step is allows us to use the decoupling trick in unquantized MIMO-OFDM.

Curiously, the EM method for iteratively decoupling the one-bit MIMO-OFDM detection problem does not seem to have caught much attention.
In this paper we aim to exploit this possibility.

\section{EM for One-Bit MIMO-OFDM Detection}

\label{sec:EM}

In the current study we focus on the box relaxation of the ML problem \eqref{eq:ml2} under the quadrature-amplitude modulation (QAM) constellation.
However, note that the idea may be expanded to some other (possibly better) detection methods---which will be future work.
Let
\[
\setS = \{s_R+\jj s_I| s_R,s_I \in \{\pm 1,\pm 3, \ldots, \pm(2D-1)\}  \}
\]
be the QAM constellation.
The box relaxation of problem \eqref{eq:ml2} is
\begin{equation}\label{eq:box_relax}
\min_{\bS} F(\bS), \quad \mbox{s.t.}~~ \bS \in \setU^{W\times K},
\end{equation}
where $\setU = \{s_R+\jj s_I| s_R,s_I \in [-2D+1,2D-1]  \}$.
Our interest lies in using EM to solve problem \eqref{eq:box_relax},
and we do it by a variational formulation.
%and we do so from a variational viewpoint (which is slightly different from the classical EM viewpoint).
Consider the following fact.

\begin{Fact}\label{fact:vari_bound}
Consider the signal model
 \[
  y = Q(r),\quad r= z+v,\quad v \sim {\cal N}(0,\sigma^2).
  \]
  Let $D_y$ be the class of all distributions with support $C_y$,
%Let $q$ be a distribution over all possible distribution class $\setD_y$ with support $C_y$,
where  $C_{1} = \Rbb_+$ and $C_{-1} =\Rbb_{-}$.
We have
  \begin{equation}\label{eq:fact1}
  \begin{split}
   -\log p(y|z) = -\log \Phi \left( \frac{yz}{\sigma}\right)
     = \min_{q\in \setD_y} ~ g(z, q)
    \end{split}
  \end{equation}
  where
  \begin{equation}\label{eq:vari_func}
    \begin{split}
%      g(z, q):=&~\frac{1}{2\sigma^2} (|\Exp_{r\sim q}[r] -z|^2+{\rm var}_{r\sim q}[r] )\\
%      &~+ \log(\sqrt{2\pi}\sigma) +\Exp_{r\sim q}[\log q(r)].
      g(z, q):=&~\frac{1}{2\sigma^2} |\Exp_{r\sim q}[r] -z|^2 +h(q),\\
     h(q):= & \frac{1}{2\sigma^2} {\rm var}_{r\sim q}[r] + \log(\sqrt{2\pi}\sigma) +\Exp_{r\sim q}[\log q(r)].
    \end{split}
  \end{equation}
  The optimal solution $q^\star$ to the problem in \eqref{eq:fact1} is a truncated Gaussian distribution
  \[
   q^{\star}(r) = \frac{{\cal N}(r;z,\sigma^2) \mathbb{1}_{C_y}(r)}{\int_{C_y}{\cal N}(r;z,\sigma^2)dr},
   \quad
  \mathbb{1}_{\setX}(x) =\begin{cases}
  0, & \mbox{if } x\notin \setX \\
  1, & \mbox{if } x \in \setX
  \end{cases}
  \]
  Also, the mean of the truncated  Gaussian distribution is given by
  \begin{equation} \label{eq:trun_mean}
   %\Exp_{r\sim q^\star}[r] = x +  \frac{1}{\sqrt{2\pi}} \frac{e^{-x^2/(2\sigma^2)}}{\Phi(-x/\sigma)}\sigma.
   \Exp_{r\sim q^{\star}}[r] = z+ y \frac{1}{\sqrt{2\pi}}\frac{e^{-z^2/(2\sigma^2)}}{\Phi( z y/\sigma)}\sigma.
  \end{equation}
\end{Fact}

Fact~\ref{fact:vari_bound} is a consequence of Jensen's inequality.
The proof can be found in  \cite{mezghani2010multiple,plabst2018efficient} (the style of presentation there is different, but the result is the same),
and it is omitted here.
By applying the variational characterization \eqref{eq:fact1} to $f_{n,w}(\bm S)$ in \eqref{eq:ml2}, we can recast the problem~\eqref{eq:box_relax} as
\begin{equation}\label{eq:vari_opt}
  \begin{split}
    \min_{\bS\in \setU^{W\times K}} \! \min_{\{q_{n,w}\}} G(\bS,\{q_{n,w}\})\! := \!\sum_{n=1}^N \!\sum_{w=1}^W g_{n,w}(z_{n,w}, q_{n,w} ),
  \end{split}
\end{equation}
where $q_{n,w} = \{q^R_{n,w}, q^I_{n,w}\}$;
$q^R_{n,w}\in \setD_{\Ree(y_{n,w})}$, $q^I_{n,w}\in \setD_{\Imm(y_{n,w})}$;
\begin{equation*}
  \begin{split}
    g_{n,w}(z_{n,w}, q_{n,w}  ) :=~ & g(\Ree(z_{n,w}), q^R_{n,w} )+ g(\Imm(z_{n,w}), q^I_{n,w} )
    %\\
    %q_{n,w} = & \{q^R_{n,w}, q^I_{n,w}\}
  \end{split}
\end{equation*}
with $g$ given by \eqref{eq:vari_func}.
%$g_{n,w}^R$ and $g_{n,w}^I$ defined taking the form in \eqref{eq:vari_func}.
Problem \eqref{eq:vari_opt} shows a structure suitable for the application of  block coordinate descent (BCD), which we will explore in the subsequent subsections.
%The two-block structure of problem \eqref{eq:vari_opt} motivates the block coordinate descent (BCD) algorithm by alternately optimizing the two blocks $\bS$ and $\{q_{n,w} \}$.
%We discuss three types of BCD algorithms, and show their merits and downsides.

\subsection{Exact BCD: EM Algorithm}
\label{sec:exact_EM}
Let us start with the classical exact BCD:
%Exact BCD performs the update rules at the $j+1$th iteration:
\begin{subequations}\label{eq:MM_exact}
  \begin{align}
  \{ q_{n,w}^{j+1} \} = & \arg\min_{\{ q_{n,w} \}} G(\bS^{j},\{q_{n,w}\}),~ \forall n,w,\label{eq:MM_exact_q}\\
  \bS^{j+1} = & \arg\min_{\bS\in \setU^{W\times K}} G(\bS,\{q_{n,w}^{j+1}\}).\label{eq:MM_exact_s}
  \end{align}
\end{subequations}
The above exact BCD scheme is an instance of the classical EM: \eqref{eq:MM_exact_q} is the E-step in EM, while \eqref{eq:MM_exact_s} the M-step.
Let us study the solutions to \eqref{eq:MM_exact_q}--\eqref{eq:MM_exact_s}.
By Fact~\ref{fact:vari_bound}, the E-step \eqref{eq:MM_exact_q} admits a closed form
%\begin{equation}\label{eq:q_R}
%    (q^R_{n,w})^{j+1} = \frac{{\cal N}(r; \Ree(z_{n,w}^{j+1}) ,\sigma^2)\mathbb{1}_{C_{\Ree(y_{n,w})}}(r)}{\int_{C_{\Ree(y_{n,w})}}{\cal N}(r;\Ree(z_{n,w}^{j+1},\sigma^2)dr}, ~\forall n,w,
%\end{equation}
\begin{align*}
(q^R_{n,w})^{j+1} & = \frac{{\cal N}(r; \Ree(z_{n,w}^{j+1}) ,\sigma^2)\mathbb{1}_{C_{\Ree(y_{n,w})}}(r)}{\int_{C_{\Ree(y_{n,w})}}{\cal N}(r;\Ree(z_{n,w}^{j+1},\sigma^2)dr}, % ~\forall n,w,
\\
(q^I_{n,w})^{j+1} & = \frac{{\cal N}(r; \Imm(z_{n,w}^{j+1}) ,\sigma^2)\mathbb{1}_{C_{\Imm(y_{n,w})}}(r)}{\int_{C_{\Imm(y_{n,w})}}{\cal N}(r;\Imm(z_{n,w}^{j+1},\sigma^2)dr},
\end{align*}
for all $n,w$, where
$\bm z_n^{j+1} = \textstyle \bm F^H ( \sum_{k=1}^K  ( \check{\bm h}_{n,k} \odot \bm s_k^{j} ))$.
To describe the M-step \eqref{eq:MM_exact_s}, let $r_{n,w}^{j+1}$ whose real part is
\begin{equation}\label{eq:E_update}
\begin{split}
&\Ree(r_{n,w}^{j+1})= \Exp_{r\sim (q^R_{n,w})^{j+1}}[r] \\
= & \Ree(z_{n,w}^{j+1})+  \Ree(y_{n,w}) \frac{1}{\sqrt{2\pi}}\frac{e^{-\Ree(z_{n,w}^{j+1})^2/(2\sigma^2)}}{\Phi( \Ree(z_{n,w}^{j+1}) \Ree(y_{n,w}) /\sigma)}\sigma;
\end{split}
\end{equation}
and whose imaginary part $\Imm(r_{n,w}^{j+1})$ is constructed by the same way as above, with ``$\Ree$'' and ``$R$'' replaced by ``$\Imm$'' and ``$I$'', respectively.
Note that, from \eqref{eq:vari_func},
\begin{equation*}
  \begin{split}
    g_{n,w}(z_{n,w}, q_{n,w}^{j+1}  ) = &~ \frac{1}{2 \sigma^2} | r_{n,w}^{j+1} - z_{n,w} |^2 +h_{n,w}(q_{n,w}^{j+1}),\\
    h_{n,w}(q_{n,w}) := & ~ h( q^R_{n,w} )+ h( q^I_{n,w} ).
  \end{split}
\end{equation*}
%where the above constant is independent of $z_{n,w}$.
By applying the decoupling trick in unquantized MIMO-OFDM in \eqref{eq:ml1}-\eqref{eq:ml3},
the M-step problem in \eqref{eq:MM_exact_s} can be rewritten as
\begin{equation}\label{eq:M_exact}
\check{\bm s}_w^{j+1}= \arg\min_{ \check{\bm s}_w \in \mathcal{U}^K } \| \check{\bm r}_w^{j+1} - \bm H_w \check{\bm s}_w \|^2, \quad w=1,\ldots,W,
\end{equation}
where we decouple the problem among subcarriers.
Each problem in \eqref{eq:M_exact}  is an instance of the box-constrained quadratic program, which can be solved by algorithms such as the projected gradient (PG) algorithm and ADMM \cite{boyd2011distributed}.

To summarize, the EM iteration \eqref{eq:MM_exact} allows us to decouple the one-bit MIMO-OFDM detection problem into subcarrier-wise independent MIMO detection problems that take the same form as classical MIMO detection.

\begin{Remark}
Let us discuss the computational complexity of the EM algorithm \eqref{eq:MM_exact}.
In the E-step \eqref{eq:MM_exact_q}, calculating the $\br_n^{j+1}$'s requires $2WN$ times of the calculation of function $\Phi$.
In the M-step \eqref{eq:MM_exact_s}, if we use PG to solve each problem \eqref{eq:M_exact}, the per-iteration complexity is ${\cal O}(NK)$; note that this process can be parallelized.
The conversions from E-step to M-step, and from M-step to E-step, require  DFT and IDFT, respectively, which demands ${\cal O}(N W\log_2 W)$ if FFT (IFFT) algorithm is used.
It is desirable that the number of conversions, or the number of EM iterations, is small.
There is a balance to strike between the accuracy of M-step and the computational burden of EM: exact M-step consumes a large amount of computations in solving \eqref{eq:M_exact}, but it may help reduce the number of EM iterations.
Note that we want to keep the number of EM iterations small as each EM iteration requires us to evaluate the E-step and DFT/IDFT conversion.
On the other hand, a not so accurate execution of the M-step saves computations in the M-step, but it may increase the number of EM iterations,
%and the number of evaluating the E-step and DFT/IDFT conversion.
and subsequently, the number of times that DFT/IDFT and $\Phi$ are called.
%We further investigate this issue in the following schemes.
We should mention that the EM algorithm based on unconstrained relaxation in \cite{plabst2018efficient} results in a closed-form update for M-step; however, its performance is inferior to the box relaxation studied in this paper, as simulation results will show.
\end{Remark}

\subsection{Inexact BCD: One-Step PG in M-Step}
\label{sec:one_step}
%A modification of the M-step is to
Consider an inexact variant of the BCD in \eqref{eq:MM_exact} where we inexactly solve the problem~\eqref{eq:M_exact} by a one-step PG method, starting from the previous iterate $\check{\bm s}_w^{j}$.
Specifically, the E-step \eqref{eq:MM_exact_q} remains unchanged, and the M-step \eqref{eq:MM_exact_s} is changed to
\begin{equation}\label{eq:M_onestep}
\check{\bm s}_w^{j+1} = \Pi_{\setU^K} (\check{\bm s}_w^{j} - 1/L_w (\bH_w^H\bH_w\check{\bm s}_w^{j} - \bH^H \check{\bm r}_w^{j+1}))
\end{equation}
for $w=1,\ldots, W$, where $\Pi_{\setU^K}(\bx)$ denotes the projection of a vector $\bx\in \Cbb^K$ onto $\setU^K$,
and  it equals
\begin{equation*}
\begin{split}
\tilde{\bx} =\Pi_{\setU^K}(\bx)
\Leftrightarrow  \tilde{x}_k =[\Ree(x_k)]_{-2D+1}^{2D-1} +\jj [\Imm(x_k)]_{-2D+1}^{2D-1},~ k=1,\ldots, K,
\end{split}
\end{equation*}
%and can be done by  element-wisely evaluating the projection $\Pi_{\setU}$, which is given by
%\[
% \textstyle \Pi_{\setU}(x) =[\Ree(x)]_{-2D+1}^{2D-1} +\jj [\Imm(x)]_{-2D+1}^{2D-1},
%\]
with $[x]_a^b := \min\{ \max\{x,a\}, b \}$;
$L_w = 2\sigma_{\max}(\bH_w)^2$ with $\sigma_{\max}(\bH_w)$ being the largest singular value of $\bH_w$.
One may be curious about the convergence of such inexact EM scheme.
The answer is yes, and the idea is to see the above inexact EM scheme as an instance of majorization-minimization \cite{razaviyayn2013unified}; we  omit the details owing to space limitation.
%It is guaranteed that this EM scheme converge to an optimal solution of problem \eqref{eq:box_relax} by invoking our recent result \cite[Theorem 3]{shao2019framework}.

It is interesting to note that this one-step PG inexact EM scheme looks similar (but is not identical) to 1BOX \cite{Mirfarshbafan2020}, where a PG algorithm is directly applied to solve problem~\eqref{eq:box_relax}.
Particularly, they have the same order of per-iteration complexity of evaluating the $\Phi$ function, DFT/IDFT conversion and matrix-vector products.
%A subtle optimization flaw of 1BOX is that the stepsize of the PG algorithm is hard or expensive to choose.
%%; line search methods of finding the stepsize requires IDFT for frequency-to-time domain conversion and thus is computationally expensive.
%Instead,  the stepsize of  \eqref{eq:M_onestep} can be explicitly found as $L_w$.
Our empirical experience suggests that both the one-step PG EM (\eqref{eq:MM_exact_q} and \eqref{eq:M_onestep}) and  1BOX suffer from slow convergence in terms of the numbers of EM iterations, and they incur a large amount of computations for evaluating DFT/IDFT.

\subsection{Inexact BCD: Inexact Update in M-Step}
\label{sec:inexact_BCD}
This subsection presents an alternative scheme that may provide a better balance between the accuracy of M-step and the number of EM iterations.
We apply FISTA-type accelerated projected gradient (APG) to problem \eqref{eq:M_exact} with limited number of iterations $B$ (e.g. 5$\sim$10 iterations) \cite{beck2009fast}.
Let  $\bar{\bx} = \check{\bm s}^{j}$ be the input of the APG solver, where we omit the subscript $w$ for notational simplicity.
Starting from $\bar{\bx}$, the APG solver in each iteration $i$,  $i=0,\ldots, B-1$, reads as
\begin{equation}\label{eq:APG}
\bm x^{i+1} = \Pi_{\setU^{K}}(\bw^{i} - 1/L (\bH^H\bH\bw^{i} - \bH^H \check{\bm r}^{j+1})),
\end{equation}
where  $\bw^i = \bx^{i}+\alpha_i (\bx^i-\bx^{i-1})$ with
$  \alpha_i = \frac{\xi_{i-1}-1}{\xi_i}, \quad \xi_i = \frac{\sqrt{1+4\xi_{i-1}^2}}{2}$,
 $\xi_{-1} = 1$ and $\bx^0 =\bx^{-1} =\bar{\bx}$.
Then, we set $\check{\bm s}^{j+1} =  \bm x^{B}$ as the output.
The APG method has a theoretical faster convergence rate than the PG method.
As of the writing of this paper, we are unable to definitively confirm whether the above inexact BCD scheme guarantees convergence to an optimal solution.
Some related studies in optimization, such as \cite{xu2017globally,wu2020hybrid}, provide strong indication that the answer should be yes.
However, the optimization variables $q_{n,w}$'s are distributions, not vectors.
This gives us new challenge.
Convergence analysis will be our future work.

We summarize the above three EM algorithms in Algorithm~\ref{alg:EM}.

\begin{algorithm}[htb!]
	\caption{EM for One-Bit MIMO-OFDM Detection} \label{alg:EM}
	\begin{algorithmic}[1]		
		\STATE {\bf given} an initialization $\bS^{0}$, iteration number $j= 0$.
%increasing $\lambda$ will make $\cal H$ as a more accurate approximation of problem \eqref{eq:ONMF_non_pen}, but more challenging to solve
		% \STATE {\bf repeat}
        \STATE {\bf repeat}

		\STATE
~~$\bm z_n^{j+1} = \textstyle \bm F^H ( \sum_{k=1}^K  ( \check{\bm h}_{n,k} \odot \bm s_k^{j} ))$
\STATE ~~{\bf E-step:}
\STATE ~~$\Ree(r_{n,w}^{j+1})\!=\!\Ree(z_{n,w}^{j+1}) \!+\!  \Ree(y_{n,w}) \frac{\sigma}{\sqrt{2\pi}}\frac{e^{-\Ree(z_{n,w}^{j+1})^2/(2\sigma^2)}}{\Phi( \Ree(z_{n,w}^{j+1}) \Ree(y_{n,w}) /\sigma)}$
\STATE ~~$\Imm(r_{n,w}^{j+1})\!=\!\Imm(z_{n,w}^{j+1}) \!+\!  \Imm(y_{n,w}) \frac{\sigma}{\sqrt{2\pi}}\frac{e^{-\Imm(z_{n,w}^{j+1})^2/(2\sigma^2)}}{\Phi( \Imm(z_{n,w}^{j+1}) \Imm(y_{n,w}) /\sigma)}$
\STATE ~~$\tilde{\bm r}_n^{j+1} = \bm F \bm r_n^{j+1}$,  $\check{\bm r}_w^{j+1} = (\tilde{r}_{1,w}^{j+1},\ldots, \tilde{r}_{N,w}^{j+1} )$
\STATE ~~{\bf M-step:}
\STATE ~~update $\bS^{j+1}$ by one of the three strategies in Sections~\ref{sec:exact_EM}- \ref{sec:inexact_BCD}
		\STATE ~~$j= j+1$
        \STATE  {\bf until} {some stopping criterion is satisfied.   }
		% \STATE {\bf until} $\lambda_k$ is larger than a pre-specified threshold		
	\end{algorithmic}
\end{algorithm}

\section{Simulation Results}

In this section, we demonstrate the performance of the EM algorithms by numerical simulations.
We consider the inexact EM, or BCD, scheme in Section~\ref{sec:inexact_BCD}, with $B=5$ APG steps for each EM iteration.
The benchmarked algorithms are the zero-forcing (ZF) detectors over each subcarrier and under full-resolution and one-bit ADCs, the 1BOX algorithm~\cite{Mirfarshbafan2020} and the unconstrained regularized EM algorithm \cite{plabst2018efficient}.
%The algorithm setting provided in \cite{Mirfarshbafan2020} does not work well under the channel models to be described below.
%We tried our best to improve the performance of 1BOX, by decreasing the stepsize from $\sqrt{2}/64$ to 1/512 and increasing the iteration number from 3 to 200; unfortunately, 1BOX does not perform well in some cases, especially when the problem size is large and when the SNR is high.
For 1BOX, we decrease the stepsize from $\sqrt{2}/64$ to 1/512 and increase the iteration number from 3 to 200 for better performance~\cite{Mirfarshbafan2020}.
We terminate our EM algorithm if the difference of two successive iterates satisfies $\textstyle \| \bS^{j+1}- \bS^{j}\|_F\leq 2\times 10^{-4}\| \bS^{j}\|_F$ or if the maximum iteration number $200$ is reached.

We adopt a widely-used multipath channel model for millimeter-wave massive MIMO with uniform linear array at the BS \cite{heath2016overview}.
The time-domain channel from user $k$ to all the antennas of the BS at the $l$th tap  is generated by
\[
  \textstyle  \bh_{l,k} = \sum_{i=1}^{\eta} \beta_{l,k}^i \ba^i_{l,k},~ l=1,\ldots,L,~k=1,\ldots,K,
\]
where $L$ is the number of channel taps; $\beta_{l,k}^i\sim \frac{1}{\sqrt{\eta}} {\cal CN}(0,1)$ is the complex channel gain; $\ba^i_{l,k}$ is the steering vector
\[
 \ba^i_{l,k}  =[1, e^{-\jj \frac{2\pi d}{\lambda}\sin(\theta^i_{l,k})}, \ldots, e^{-\jj \frac{2\pi d}{\lambda}(N-1)\sin(\theta^i_{l,k})}]^T,
\]
with $d$, $\lambda$ and $\theta^i_{l,k}$ being the inter-antenna spacing, carrier wavelength and path angle, respectively; $\eta$ is the number of paths.
Throughout the simulation, we set $L=16$, $\frac{d}{\lambda}=\frac{1}{2}$, $\eta=4$ and randomly generate the path angles $\theta^i_{l,k}$'s from $[-\pi,\pi]$.
% The number of channel taps is set as $L=16$.
%The other simulation settings are as follows.
%The number of antennas is $N=256$ and the number of users is $K=18$.
%The number  of OFDM subcarriers is $W=512$.

%\begin{figure}[htb!]
%  \centering
%  \includegraphics[width=0.65\linewidth]{./fig/BER_4QAM_W512_N256_K36.eps}
%  \caption{BER performance. $(N,K,W) = (256,36, 512)$, $4$-ary QAM.}\label{fig:BER_4QAM}
%\end{figure}
%
%
%\begin{figure}[htb!]
%  \centering
%  \includegraphics[width=0.65\linewidth]{./fig/BER_16QAM_W512_N256_K20.eps}
%  \caption{BER performance. $(N,K,W) = (256,20, 512)$, $16$-ary QAM.}\label{fig:ber}
%\end{figure}
\begin{figure}[htb!]
  \centering
  \begin{subfigure}[b]{0.495\linewidth}
  \includegraphics[width=\linewidth]{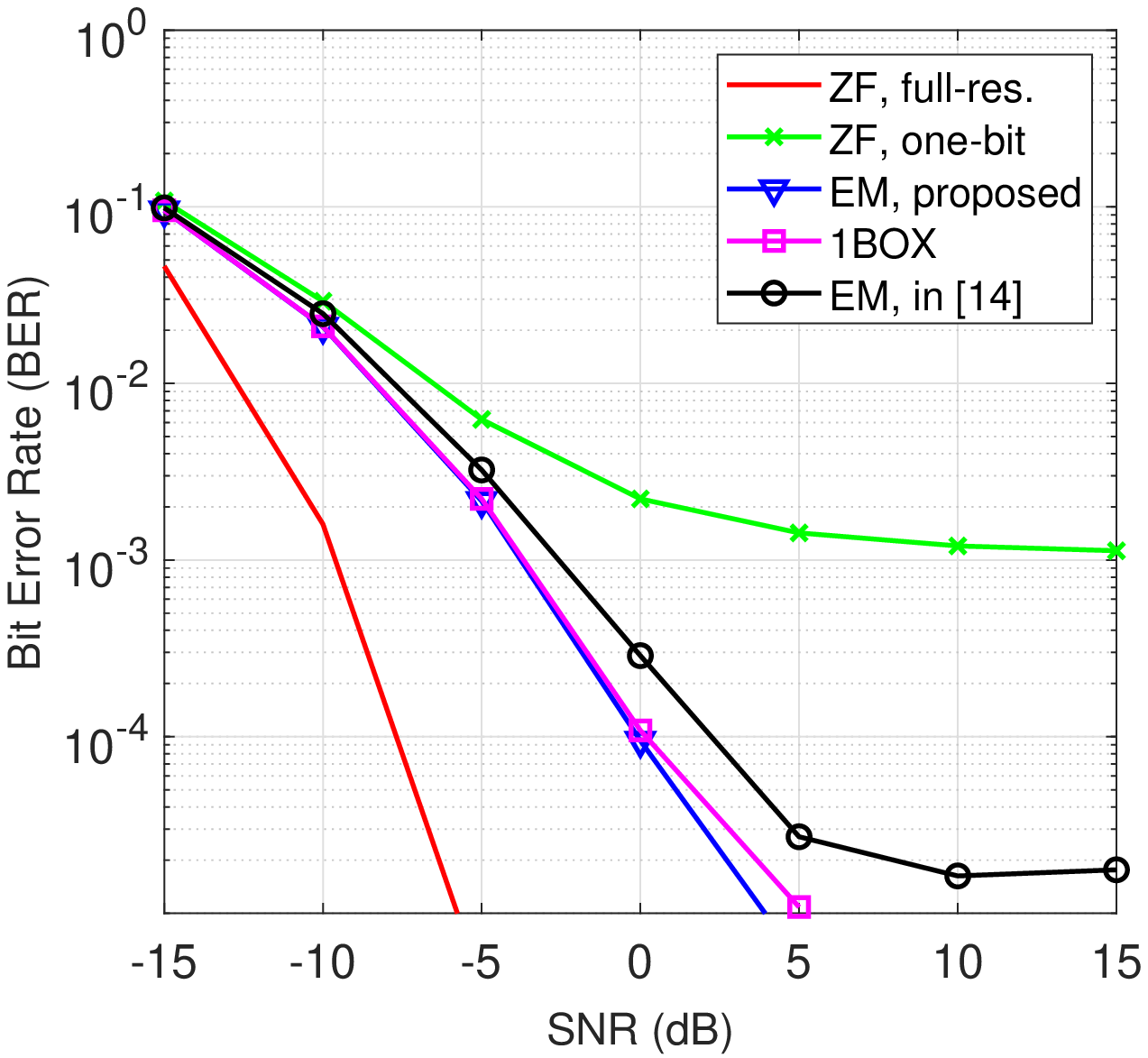}
  \caption{$(N,K,W) = (256,36, 512)$, $4$-ary QAM.}\label{fig:ber_4QAM}
  \end{subfigure}
   \begin{subfigure}[b]{0.495\linewidth}
  \includegraphics[width=\linewidth]{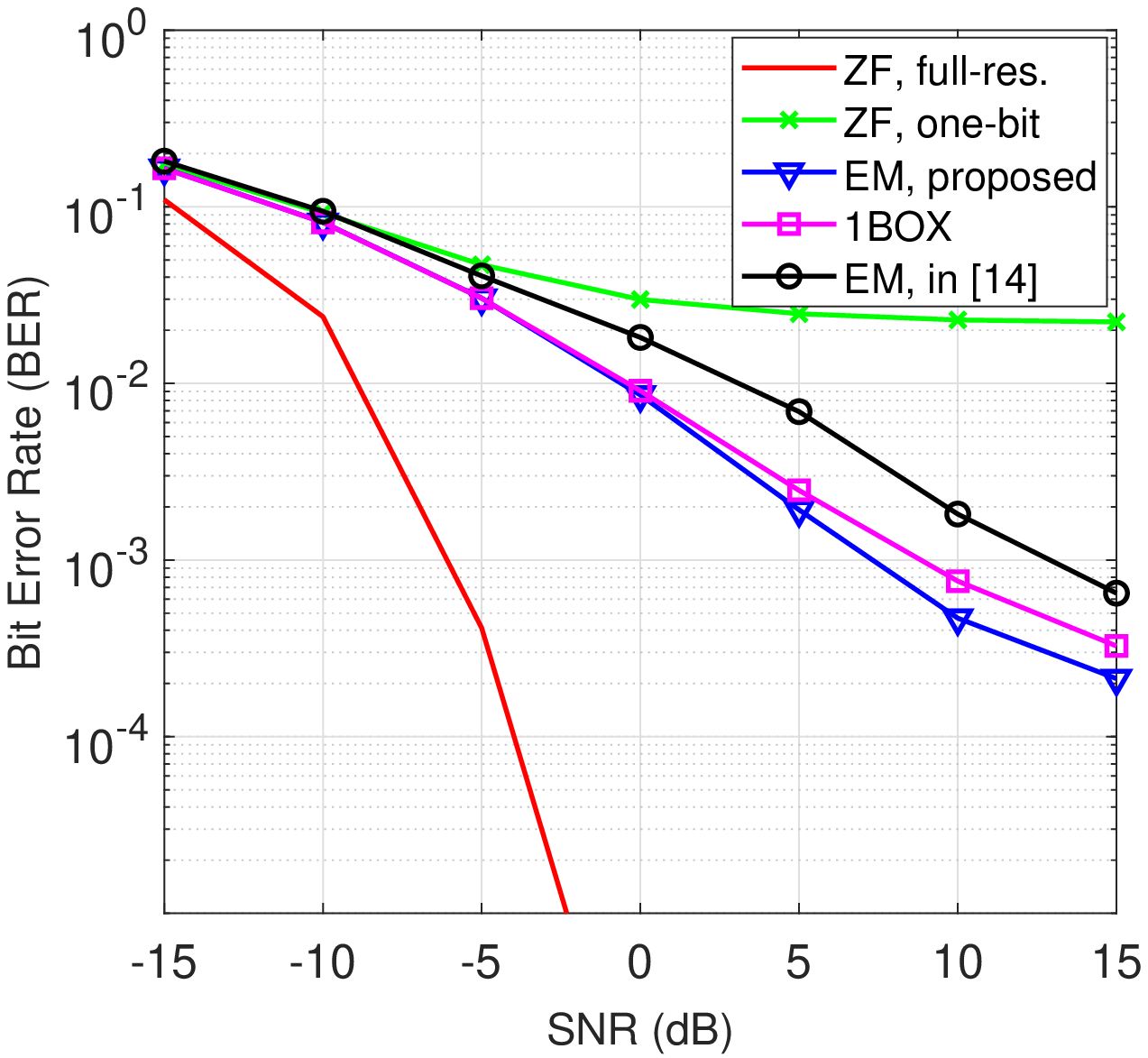}
  \caption{$(N,K,W) = (256,20, 512)$, $16$-ary QAM.}\label{fig:ber_16QAM}
  \end{subfigure}
  \caption{BER performance. }\label{fig:ber}
\end{figure}

\begin{figure}[htb!]
  \centering
  \includegraphics[width=0.7\linewidth]{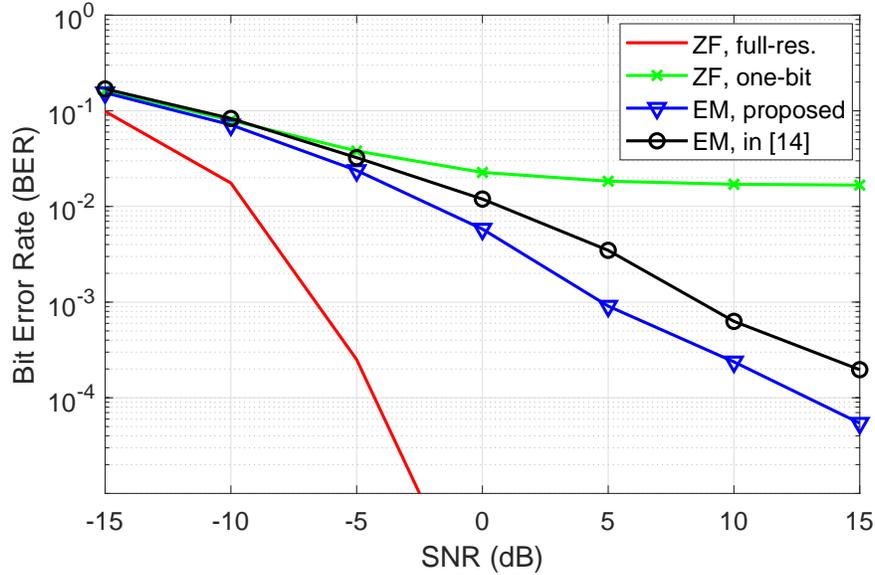}
  \caption{BER\! performance.\! $(\!N,\!K,\!W) \! =   \! (512,\!36,\! 1024)$,\! $16$-ary \! QAM.}\label{fig:BER_large}
\end{figure}

Fig.~\ref{fig:ber} shows the bit-error rate (BER) performance under 4-ary and 16-ary QAM  constellations.
The number of antennas, the number of users and the number  of OFDM subcarriers are $(N,K,W) = (256,36, 512)$ in Fig.~\ref{fig:ber}(a) and $(N,K,W) = (256,20, 512)$ in Fig.~\ref{fig:ber}(b).
We see that the EM algorithms and 1BOX achieve better performance than the one-bit ZF detector.
% except that 1BOX suffers from error floor in high SNR region.
Also, it is seen that the proposed EM algorithm performs better than the unconstrained regularized EM in \cite{plabst2018efficient}; the performance gain at the mid-to-high SNR region is about $5$ dB.
In Fig.~\ref{fig:BER_large}, we increase the problem dimension to $(N,K,W) = (512,32, 2048)$ and test the algorithms under $16$-ary QAM. We were unsuccessful in obtaining reasonable results with 1BOX.
Encouragingly, it is seen that the EM algorithms can still yield satisfactory BER performance.
Again, the proposed EM algorithm shows better BER performance than the  EM in \cite{plabst2018efficient}.

%
%\begin{figure}[htb!]
%  \centering
%  \begin{subfigure}[b]{0.65\linewidth}
%  \includegraphics[width=\linewidth]{./fig/BER_4QAM_W512_N256_K36.eps}
%  \caption{$(N,K,W) = (256,36, 512)$, $4$-ary QAM.}\label{fig:ber_4QAM}
%  \end{subfigure}
%   \begin{subfigure}[b]{0.65\linewidth}
%  \includegraphics[width=\linewidth]{./fig/BER_16QAM_W512_N256_K20.eps}
%  \caption{$(N,K,W) = (256,20, 512)$, $16$-ary QAM.}\label{fig:ber_16QAM}
%  \end{subfigure}
%  \caption{BER performance. }\label{fig:ber}
%\end{figure}

%
%\begin{figure}[htb!]
%  \centering
%  \begin{subfigure}[b]{0.65\linewidth}
%  \includegraphics[width=\linewidth]{./fig/onebit_OFDM_QAM_256_18_512.eps}
%  \caption{$16$-ary QAM}\label{fig:ber_16QAM}
%  \end{subfigure}
%   \begin{subfigure}[b]{0.65\linewidth}
%  \includegraphics[width=\linewidth]{./fig/onebit_OFDM_PSK_256_18_512.eps}
%  \caption{$8$-ary PSK}\label{fig:ber_8PSK}
%  \end{subfigure}
%  \caption{BER performance. $(N,K,W) = (256,18, 512)$.}\label{fig:ber}
%\end{figure}

\begin{figure}[htb!]
  \centering
  \includegraphics[width=0.7\linewidth]{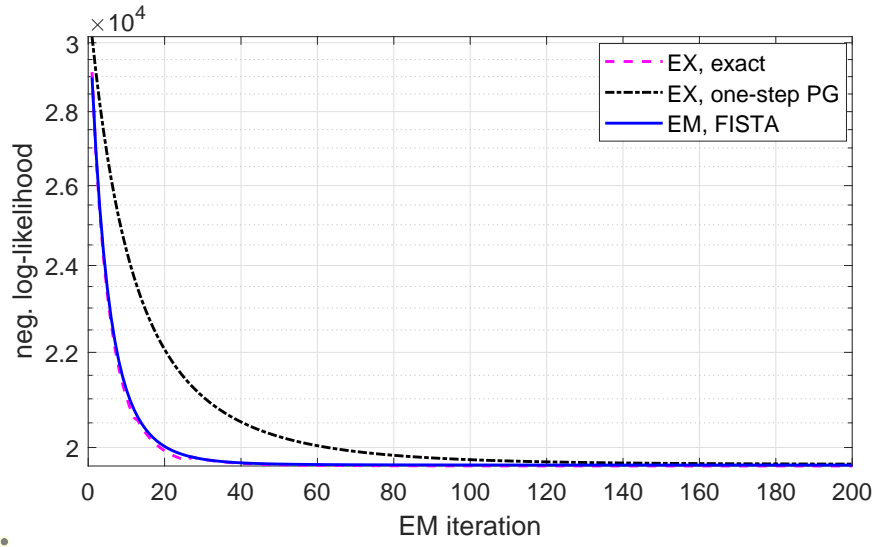}
  \caption{Convergence of EM algorithms.}\label{fig:conv}
\end{figure}

In Fig.~\ref{fig:conv}, we show the convergence of EM algorithms  in Section~\ref{sec:EM}, namely, 1) EM by exact BCD, 2) EM with one-step PG in the M-step,
3) EM with $B=5$ APG steps in the M-step,
 by showing the negative log-likelihood value $f(\bS)$ in problem \eqref{eq:ml2} versus the iteration number in a random trial.
The setting is the same as those in Fig.~\ref{fig:ber}(a) with SNR = 5~dB.
We see that the negative log-likelihood monotonically decreases with respect to EM iterations for all the EM algorithms.
Also, the inexact FISTA-type EM achieves almost the same convergence rate as the exact EM, which converges in $20\sim 40$ EM iterations.

\section{Conclusion}

In this paper, we studied exact and inexact EM schemes for one-bit ML MIMO-OFDM detection.
We demonstrated the efficiency of handling one-bit MIMO-OFDM detection
by the divide-and-conquer strategy enabled by EM.
Future work will study how this divide-and-conquer strategy can be expanded to other MIMO detection methods, particularly those that have powerful detection performance in classical MIMO detection.

%by revisiting the unquantized MIMO-OFDM detection, and it laid a foundation for exploring future directions, such as dealing with multiple-bit ADCs.
%The salient feature of the EM method is to orthogonally separate the problem among subcarriers and solve them by feat of luxuriant classical ML MIMO detection methods in the M-step, followed by a  closed-form E-step.
%The EM method allows us to revisit the rich seam of classical MIMO detection research for handling the one-bit MIMO-OFDM detection problem.
%Numerous possible future directions , including  dealing with multiple-bit ADCs, and channel estimation.

% -------------------------------------------------------------------------

%\vfill\pagebreak
%\newpage

% \bibliographystyle{IEEEbib}
\bibliographystyle{IEEEtran}
\bibliography{refs}
%\nocite{*}

\end{document}